\documentclass[twocolumn,floats,showpacs,superscriptaddress,pre]{revtex4}
\usepackage{amsmath,amssymb,bm}
\usepackage{graphics}
\usepackage{epsfig}
\usepackage{graphicx}

\begin{document}

\title{On a Probabilistic Interpretation of Relativistic Quantum Mechanics}
\author{Natalia Gorobey, Alexander Lukyanenko}
\email{alex.lukyan@rambler.ru}
\affiliation{Department of Experimental Physics, St. Petersburg State Polytechnical
University, Polytekhnicheskaya 29, 195251, St. Petersburg, Russia}
\author{Inna Lukyanenko}
\email{inna.lukyanen@gmail.com}
\affiliation{Institut f\"{u}r Mathematik, TU Berlin, Strasse des 17 Juni 136, 10623
Berlin, Germany}

\begin{abstract}
A probabilistic interpretation of one-particle relativistic quantum
mechanics is proposed. Quantum Action Principle formulated earlier is used
for to make the dynamics of the Minkowsky time variable of a particle to be
classical. After that, quantum dynamics of a particle in the $3D$ space
obtains the ordinary probabilistic interpretation. In addition, the
classical dynamics of the Minkowsky time variable may serve as a tool for
"observation" of the quantum dynamics of a particle. A relativistic analog of
the hydrogen atom energy spectrum is obtained.
\end{abstract}

\maketitle
\date{\today }





\section{\textbf{INTRODUCTION}}

The subject of the present work is the problem of a probabilistic
interpretation of relativistic quantum mechanics (RQM) based on the
Klein-Gordon (KG) wave equation for a particle in an external
electromagnetic field (see, for example, \cite{BD}),
\begin{equation}
\widehat{F}\psi \equiv \left[ \left( \frac{\hbar }{i}\nabla _{\mu }-eA_{\mu
}\right) \left( \frac{\hbar }{i}\nabla ^{\mu }-eA^{\mu }\right) -m^{2}c^{2}%
\right] \psi =0.  \label{1}
\end{equation}%
In the present paper we develope an approach that may be called "refreshed" dynamics in
contrast to the term "frosen" dynamics, which is used in quantum theory of
the universe based on the Wheeler-De Witt wave equation \cite{MTW}. The
"refreshment" of the one-particle quantum dynamics will be performed by use
of a quantum action principle (QAP) \cite{GLL1} which was applied first
to the problem of probabilistic interpretation of RQM in \cite{GLL2}. In QAP
the KG equation (\ref{1}) is replaced by the Schr\"{o}dinger equation,
\begin{equation}
i\hbar \frac{\partial \psi }{\partial s}=\widehat{F}\psi ,  \label{2}
\end{equation}%
were $s\in \left[ 0,S\right] $ is an internal time parameter of a particle.
The time interval length $S$ is an additional dynamical variable of a particle
which is not observable. It has to be excluded by use of a condition of the
stationarity of a quantum action with respect to small variations of $S$. It
is this condition of the stationarity that we call QAP. The quantum action is
defined as the real phase of a transition amplitude of a particle in a real
experiment. In the present work QAP is supplied by some more conditions of
the stationarity, which makes the dynamics of the Minkowsky time dynamical
variable $x^{0}=ct$ to be classical. In addition to the "ordinary" probabilistic
interpretation of RQM, the classical dynamics of the Minkowsky time gives us
a tool to "observe" the quantum dynamics of a particle "from the inside", and
develops its alternative interpretation. The latter will be useful in
quantum cosmology.

\section{QUANTUM ACTION PRINCIPLE IN RELATIVISTIC MECHANICS}

Schr\"{o}dinger equation (\ref{2}) arises in the result of the standard
quantization procedure applied to the classical action with an invariant time
parameter \cite{F},
\begin{equation}
I=\int\limits_{0}^{S}ds\left\{ p_{\mu }\overset{\cdot }{x}^{\mu }-\left[
\left( p_{\mu }-eA_{\mu }\right) \left( p^{\mu }-eA^{\mu }\right) -m^{2}c^{2}%
\right] \right\} .  \label{3}
\end{equation}%
The upper limit of the integration has to be defined from the condition of
the stationarity of (\ref{3}) after substitution into it a solution of classical
equations of motion. It is proportional to a proper time. For a free
moving particle it equals
\begin{equation}
S=\frac{\sqrt{\left( x_{1}-x_{0}\right) ^{2}}}{2mc},  \label{4}
\end{equation}%
were $x_{0,1}^{\mu }$ are the end points of a world line of a particle. The
main idea of the work \cite{GLL2} was that to "delay" the condition of
stationarity up to the quantum level. In the present work we slightly
"improve" the action (\ref{3}) enlarging the set of dynamical variables as
follows:
\begin{eqnarray}
\widetilde{I} &=&\int\limits_{0}^{S}ds\left\{ p_{\mu }\overset{\cdot }{x}%
^{\mu }-\left[ d^{2}-\left( p_{i}-eA_{i}\right) ^{2}-m^{2}c^{2}\right]
\right.  \notag \\
&&\left. +\lambda \left[ d-\left( p_{0}-eA_{0}\right) \right] \right\} .
\label{5}
\end{eqnarray}%
Two additional variables, $d$ and $\lambda $, may be excluded at the
classical level, and in the result the action (\ref{5}) comes back to the
form (\ref{3}). But, once again, we delay this exclusion up to the quantum
level. A result of this delay will be a classical character of the dynamics
of the Minkowsky time parameter $x^{0}\left( s\right) $, which, in turn, "refreshes" the dynamics of particle.

Let us write the Schr\"{o}dinger equation corresponding to the action (\ref{5}),
\begin{eqnarray}
i\hbar \frac{\partial \psi }{\partial s} &=&\left[ d^{2}\left( s\right)
-\lambda \left( s\right) \left( d\left( s\right) +i\hbar \frac{\partial }{%
\partial x^{0}}-k\frac{e^{2}}{cr}\right) \right.  \notag \\
&&\left. -m^{2}c^{2}+\hbar ^{2}\Delta \right] \psi ,  \label{6}
\end{eqnarray}%
where for simplicity we consider an electron movement in the Coulomb field
of a hydrogen atom. In this case the dependence of a solution $\psi $ on the
variable $x^{0}$ may be considered separately. One write $\psi \left( s,x^{\mu
}\right) =\psi _{0}\left( s,x^{0}\right) \varphi \left( s,x^{i}\right) $,
where
\begin{equation}
\psi _{0}\left( s,x^{0}\right) =\exp \chi \left( s,x^{0}\right) .  \label{7}
\end{equation}%
Here $\chi \left( s,x^{0}\right) $ is a complex phase for which a
quadratic representation
\begin{equation}
\chi \left( s,x^{0}\right) =\chi _{0}\left( s\right) +\chi _{1}\left(
s\right) x^{0}+\frac{1}{2}\chi _{2}\left( s\right) \left( x^{0}\right) ^{2}
\label{8}
\end{equation}%
is sufficient. From the equation (\ref{6}) we obtain a set of ordinary
differential equations for coefficients $\chi _{0,1,2}$,
\begin{eqnarray}
i\hbar \overset{\cdot }{\chi _{0}} &=&d^{2}-\lambda d-m^{2}c^{2}+i\hbar
\lambda \chi _{1},  \notag \\
\overset{\cdot }{i\hbar \chi }_{1} &=&i\hbar \lambda \chi _{2},  \notag \\
\overset{\cdot }{\chi }_{2} &=&0.  \label{9}
\end{eqnarray}%
We consider the quadratic representation (\ref{8}) according
to the Gauss form of the initial wave function:
\begin{equation}
\psi _{0}\left( 0,x^{0}\right) =A\exp \left( -\frac{\left( x^{0}\right) ^{2}%
}{4\sigma ^{2}}\right) .  \label{10}
\end{equation}%
A solution of the set (\ref{9}) corresponding to the initial wave
function (\ref{10}) is
\begin{eqnarray}
\chi _{2}\left( s\right) &=&-\frac{1}{2\sigma ^{2}},  \label{11} \\
\chi _{1}\left( s\right) &=&\frac{1}{2\sigma ^{2}}\int\limits_{0}^{s}ds^{%
\prime }\lambda \left( s^{\prime }\right) ,  \notag \\
\chi _{0}\left( s\right) &=&\frac{1}{i\hbar }\int\limits_{0}^{s}ds^{\prime
}\left( d^{2}-\lambda d-m^{2}c^{2}\right)  \notag \\
&&-\frac{1}{2\sigma ^{2}}\int\limits_{0}^{s}ds^{\prime }\lambda \left(
s^{\prime }\right) \int\limits_{0}^{s^{\prime }}ds^{\prime \prime }\lambda
\left( s^{\prime \prime }\right) .  \notag
\end{eqnarray}

To complete the formulation of QAP one needs a remaining part of the phase of
a full wave function $\psi \left( s,x^{\mu }\right) $. The remaining part $%
\varphi \left( s,x^{i}\right) $ of the wave function obeys the Schr\"{o}%
dinger equation,
\begin{equation}
i\hbar \frac{\partial \varphi }{\partial s}=\widehat{H}\left( \lambda \left(
s\right) \right) \varphi \equiv -\left[ -\hbar ^{2}\Delta -k\lambda \left(
s\right) \frac{e^{2}}{cr}\right] \varphi .  \label{12}
\end{equation}%
In order to examine QAP before its application for the problem of the
probabilistic interpretation let us finish the section by consideration of a
stationary solution of the equation (\ref{12}) for a hydrogen atom.
Supposing $\lambda =const$, we consider the stationary Schr\"{o}dinger equation,
\begin{equation}
\widehat{H}\varphi =\epsilon \varphi .  \label{13}
\end{equation}%
The equation (\ref{13}) has the set of bound eigenstates with the eigenvalues
\begin{equation}
\epsilon _{n}=-\frac{\lambda ^{2}}{2mc^{2}}E_{n},\,\,\,\,\,\, E_{n}\equiv -\frac{\hbar R}{%
n^{2}},  \label{14}
\end{equation}%
where $R$ is the Rydberg constant, and $n=1,2,...$ is the principal quantum
number, and corresponding eigenfunctions are
\begin{equation}
\varphi _{n}\left( \frac{2mc}{\lambda }x^{i}\right) ,  \label{15}
\end{equation}%
where $\varphi _{n}\left( x^{i}\right) $ are ordinary non-relativistic
stationary wave functions of an electron in the hydrogen atom. In this
stationary case the remaining part of the full phase equals $\left( 1/i\hbar
\right) \epsilon _{n}s$ , and the real part of the phase is
\begin{equation}
\left( d^{2}-\lambda d-m^{2}c^{2}-\frac{\lambda ^{2}}{2mc^{2}}E_{n}\right) s.
\label{16}
\end{equation}%
One can see from the representation (\ref{8}) and the solution (\ref{11})
for its coefficients that
\begin{eqnarray}
\psi _{0}\left( s,x^{0}\right) &=&A\exp \left[ -\frac{\left( x^{0}-\lambda
s\right) ^{2}}{4\sigma ^{2}}\right.  \label{17} \\
&&\left. -\frac{i}{\hbar }\left( d^{2}-\lambda d-m^{2}c^{2}-\frac{\lambda
^{2}}{2mc^{2}}E_{n}\right) s\right] .  \notag
\end{eqnarray}%
It means that $\lambda S$ equals to the end value of the Minkowsky time
coordinate $x_{1}^{0}$. The quantum action defined as the real phase (\ref%
{16}) with account of this additional condition is
\begin{eqnarray}
\Lambda &=&\left( d^{2}-\lambda d-m^{2}c^{2}-\frac{\lambda ^{2}}{2mc^{2}}%
E_{n}\right) S  \label{18} \\
&&+\varkappa \left( \lambda S-x_{1}^{0}\right) ,  \notag
\end{eqnarray}%
where $\varkappa $ is a corresponding Lagrangian multiplier. We formulate QAP
as a set of conditions of the stationarity of the action (\ref{18}) with respect
to small variations of the parameters $d,\lambda ,S,\varkappa $. Their
stationary values are
\begin{eqnarray}
\lambda &=&2d=2mc/\sqrt{1+\frac{2E_{n}}{mc^{2}}},S=\frac{x_{1}^{0}}{\lambda }%
,  \notag \\
\varkappa c &=&\sqrt{m^{2}c^{4}+2mc^{2}E_{n}}.  \label{19}
\end{eqnarray}%
Compare the second equation with the classical result in the Eq. (\ref{4}).
It is the last quantity in (\ref{19}) that equals to the energy of an
electron in the hydrogen atom in our theory (as the coefficient in front of $%
t=x^{0}/c$). This result coincides with the prediction of the Dirac equation
with accuracy up to the square of the fine structure constant, $\alpha
^{2}$. In the precise energy spectrum the square of the principal quantum
number $n^{2}$ must be replaced by the following quantity \cite{Fock}:
\begin{eqnarray}
n^{\ast 2} &=&p^{2}+2p\sqrt{k^{2}-\alpha ^{2}}+k^{2},  \label{20} \\
p &=&0,1,...,\left\vert k\right\vert =1,2,....  \notag
\end{eqnarray}%
The origin of the difference is the absence of spin in our theory.

\section{PROBABILISTIC INTERPRETATION OF RELATIVISTIC QUANTUM MECHANICS}

Let us turn to dynamics. Let $\lambda \left( s\right) $ be an arbitrary
function of the internal time, and $\widehat{U}_{S}\left[ \lambda \left(
s\right) \right] $ be the evolution operator for the Schr\"{o}dinger
equation (\ref{12}) on the time interval $\left[ 0,S\right] $. For a pair of
normalized states $\left\vert \varphi _{in}\right\rangle $ and $\left\vert
\varphi _{out}\right\rangle $ one can introduce a transition amplitude,
\begin{equation}
K_{S}\left[ \lambda \left( s\right) \right] \equiv \left\langle \varphi
_{out}\right\vert \widehat{U}_{S}\left[ \lambda \left( s\right) \right]
\left\vert \varphi _{in}\right\rangle .  \label{21}
\end{equation}%
In the work \cite{GLL2} we defined a quantum action as the real phase of the
transition amplitude for a concrete experiment. Here the transition
amplitude (\ref{21}) written in the exponential form,
\begin{equation}
K_{S}\left[ \lambda \left( s\right) \right] =\exp \left\{ \frac{1}{i\hbar }I%
\left[ \lambda \left( s\right) \right] +Q\left[ \lambda \left( s\right) %
\right] \right\} ,  \label{22}
\end{equation}%
defines a part of a full quantum action that, instead of the Eq.(\ref{18}),
equals to
\begin{equation}
\Lambda =-\int\limits_{0}^{S}ds\left[ \frac{\lambda ^{2}}{4}+m^{2}c^{2}%
\right] +\varkappa \left( \int\limits_{0}^{S}ds\lambda -x_{1}^{0}\right) +I%
\left[ \lambda \left( s\right) \right] .  \label{23}
\end{equation}%
Let $\widetilde{\lambda }\left( s\right) $ be a stationary value of the
function $\lambda \left( s\right) $ on the time interval $\left[ 0,%
\widetilde{S}\right] $ with a stationary length $\widetilde{S}$. Then the
transition amplitude $K_{\widetilde{S}}\left[ \widetilde{\lambda }\left(
s\right) \right] $ becomes a complex amplitude of a probability, namely, the
probability of the quantum transition $\left\vert \varphi _{in}\right\rangle
\rightarrow \left\vert \varphi _{out}\right\rangle $ at the moment $%
x_{1}^{0} $ of the Minkowsky time equals
\begin{equation}
P_{x_{1}^{0}}=\left\vert K_{\widetilde{S}}\left[ \widetilde{\lambda }\left(
s\right) \right] \right\vert ^{2}.  \label{24}
\end{equation}

This interpretation does not depend on the dynamics of the Minkowsky time
variable $x^{0}\left( s\right) $ which in fact is classical one. The latter
follows from the Gauss form dependence on $x^{0}$ of the wave function (\ref%
{17}). In the limit $\sigma \rightarrow 0$ it becomes proportional to the
corresponding $\delta $-function. But the classical dynamics of the time
variable $x^{0}\left( s\right) $ may be considered as a tool for observation
of quantum dynamics of a particle in the $3D$ space. In a general case, as a consequence of the classical dynamics of $x^{0}\left( s\right) $, we have
\begin{equation}
\widetilde{\lambda }\left( s\right) =\frac{dx^{0}\left( s\right) }{ds}.
\label{25}
\end{equation}%
For a fixed initial state $\left\vert \varphi _{in}\right\rangle $ and
an arbitrary final state $\left\vert \varphi _{out}\right\rangle $, QAP gives a
certain stationary function $\widetilde{\lambda }\left( s\right) $ and
corresponding evolution of the internal time,
\begin{equation}
s\left( x^{0}\right) =\int\limits_{0}^{x^{0}}\frac{dx^{0}}{\widetilde{%
\lambda }\left( s\left( x^{0}\right) \right) }.  \label{26}
\end{equation}%
We suppose that the equation (\ref{26}) gives a one-to one correspondence
between the quantum transition $\left\vert \varphi _{in}\right\rangle
\rightarrow \left\vert \varphi _{out}\right\rangle $ and the evolution of
the internal time. Let us imagine for a moment a clock connected with a
quantum electron. The movement of the clock reflects the quantum dynamics of
the electron. Of course, this speculation is unreal. But in quantum
cosmology just this situation is realized: all observers with devises are
located in a quantum universe. We need in this case an infinite set of
classical degrees of freedom. Their relative dynamics can be made observable.

\section{\textbf{CONCLUSIONS }}

Therefore, QAP gives a possibility of the ordinary probabilistic interpretation
of relativistic quantum mechanics, under the condition that the dynamics of the
Minkowsky time variable of particle is classical one. The latter may be
achieved by corresponding modification of QAP. This approach may be useful
in quantum cosmology as a method of "refreshment" of quantum dynamics of the
universe.

We are thanks V. A. Franke and A. V. Goltsev for useful discussions.




\end{document}